\documentclass{aa}  
\usepackage{graphicx}
\usepackage{printlen}
\usepackage{physics}
\usepackage{siunitx}
\usepackage{caption}
\usepackage{txfonts}
\usepackage{hyperref}
\usepackage{natbib}
\bibpunct{(}{)}{;}{a}{}{,} % to follow the A&A style
\begin{document}

   \title{Effects of limited core rejuvenation on the properties of massive contact binaries}

   \author{J.~Vandersnickt\inst{\ref{inst:kul}} \and M.~Fabry\inst{\ref{inst:kul},\ref{inst:fabry}}}

    \institute{Institute of Astronomy, KU Leuven, Celestijnenlaan 200D, 3001 Leuven, Belgium \\ \email{jelle.vandersnickt@student.kuleuven.be} \label{inst:kul} \and Department of Astrophysics and Planetary Science, 800 E Lancaster Ave., PA 19085, USA\label{inst:fabry}}

   \date{}

% \abstract{}{}{}{}{} 
% 5 {} token are mandatory
 
  \abstract
   {Massive contact binaries are both stellar-merger and gravitational-wave progenitors, but their evolution is still uncertain. An open problem in the population synthesis of massive contact binaries is the predicted mass ratio distribution. Current simulations evolve quickly to mass ratios close to unity, which is not supported by the sample of observed systems. It has been shown that modifying the near core mixing properties of massive stars can alter the evolution of contact binaries, but this has not been tested on a whole population.}
   {We implement a prescription of the convective core overshooting based on the molecular gradient. The goal of the implementation is to limit the rejuvenation efficiency of the accretor. We aim to investigate the effects of the reduced rejuvenation on the mass ratio distribution of massive contact binaries. }
   {We calculated a grid of \num{4896} models using the binary-evolution code \texttt{MESA} with our convective overshoot formulation, and we compared the simulations with the known observations of massive contact binaries.   }
   {We find that by limiting core rejuvenation through the convective core overshoot, the predicted mass ratio distribution shifts significantly to values away from unity. This improves the theoretical predictions of the mass ratios of massive contact binaries. }
   {The core rejuvenation of the components in massive contact binaries is a key parameter for their predicted mass ratio distribution. Establishing the rejuvenation efficiency within the contact binary population should therefore be possible. The sample size and uncertainties associated with the characterization of contact binaries, however, prevents us from doing so, and other methods like asteroseismology can place constraints on the rejuvenation process.}


   \maketitle
\section{Introduction}
Observations indicate that the majority of massive stars ($M~\gtrsim~8$~$M_\odot$) are in binary systems \citep{2014ApJS..215...15S, 2021MNRAS.507.5348V, 2022A&A...658A..69B}. 
\citet{2012Sci...337..444S} show that half of all massive stars interact before the end of the main sequence; this is called Case A evolution \citep{1967ZA.....66...58K, 1994A&A...290..119P, 1998NewA....3..443V, 2001ApJ...552..664N}. 

Case A systems have initial orbital periods of around ten days or less. 
When such a massive binary  evolves, an extreme form of interaction can occur when both stars overflow their Roche lobes simultaneously.
This is the contact phase, and the system is then classified as a contact binary\footnote{We make no distinction between  the term `contact' and `overcontact'.}.
Previous studies show that a contact phase is expected for at least 40\% of all massive binaries with a mass-transfer event \citep{2024A&A...682A.169H}.

Massive contact binaries are likely to merge \citep{2021MNRAS.507.5013M, 2024A&A...682A.169H, Fabry2024}, which has been identified as a possible process for creating magnetic massive stars \citep{2016MNRAS.457.2355S, 2019Natur.574..211S}. 
Recently, this scenario has been observationally confirmed in HD~148937 \citep{2024Sci...384..214F}. 
Mergers are also thought to create rapidly rotating neutron stars, long gamma-ray bursts \citep{2005A&A...443..643Y, 2006A&A...460..199Y}, and super-luminous supernovae \citep{2018ApJ...858..115A, 2020ApJ...901..114A}. 
At low metallicity, massive contact binaries are expected to make a dominating contribution to the binary black hole population that is observed through gravitational waves \citep{2016MNRAS.458.2634M, 2016A&A...588A..50M, 2022PhR...955....1M}.

Despite many efforts to find massive contact binaries, only around two dozen are known.
There are several systems in the Galactic environment \citep[e.g.][]{1985MNRAS.213...75H, 2017A&A...606A..54L, 2021A&A...651A..96A, 2022ApJ...932...14L}. 
Additionally, large surveys including the Massive Compact Halo Objects (MACHO) Project \citep{1997AJ....114..326A} and the Tarantula Massive Binary Monitoring \citep{2015ApJ...812..102A, 2017A&A...598A..84A, 2020A&A...634A.119M, 2020A&A...634A.118M} detected such systems. 
These systems are well studied. 
For example, recent studies investigated the orbital variations of a sample of O- and B-type massive contact binaries and studied the evolutionary timescales of these systems \citep{2022A&A...666A..18A, 2024A&A...691A.150V}.
A more extensive overview of observed massive contact binaries can be found in Appendix \ref{sec:observations}.

The theoretical description of contact systems was initiated by \citet{1941ApJ....93..133K}, who identified that contact binaries cannot be stable if the components obey a single-star internal structure, except at a mass ratio of unity. 
Contact binaries are preferentially observed at mass ratios away from unity, however (especially in the low-mass W UMa regime), which led to Kuiper's argument being called Kuiper's paradox.
The solution to the paradox is to consider stellar models that are affected by binary interaction, for example through mass and energy transfer, as well as tidal distortion.

\citet{1968ApJ...151.1123L} was the first to  consider energy transfer within the contact binary as a possible solution. 
Other models followed these works \citep[e.g.][]{1972A&A....16...60B, 1976ApJ...205..217F, 1976ApJ...209..536S, 1977ApJ...215..851W}. However, some were found to contain inconsistencies \citep{1979MNRAS.189P...5P, 1980A&A....84..200H}, or focused only on those low-mass contact binaries with convective envelopes. 

In addition to analytical models, computational studies of contact binaries can be performed. 
This was done with 1-dimensional binary evolution codes for example by \citet{1994A&A...290..119P}, \citet{2001A&A...369..939W}, \citet{2001ApJ...552..664N}, \citet{2007IAUS..240..384D}, and \citet{2016A&A...588A..50M} for massive stars. 
More recently, the population synthesis calculation by \citet{2021MNRAS.507.5013M} showed an overestimation of equal mass ratio using current mass transfer models.
They found that unequal mass ratio systems equalize on the nuclear timescale, but the models still show that the time spent as equal mass contact binaries dominates the population. However, \citet{2023A&A...672A.175F} shows that, under certain conditions, the inclusion of energy transfer can delay the equalization of the mass ratio.
The models also account for the tidal deformation of the components \citep{2022A&A...661A.123F}. 
Despite these improvements, \citet{Fabry2024} shows that this only has a limited effect on the overall properties of the contact binary population. 
Therefore, other processes in stellar and binary physics should be explored to tackle this discrepancy.

Mass transfer significantly influences the evolution of the stars in a binary system. 
As a main sequence accretor grows during mass transfer, its core grows in parallel and rejuvenation of the accretor can take place. 
Due to the strong mixing induced by convection in massive stars, additional hydrogen gets mixed in and the hydrogen mass fraction ($X)$ increases. 
The star will look younger than predicted by single stellar evolution \citep{2007MNRAS.376...61D}. 
\citet{2016MNRAS.457.2355S} found that rejuvenation can make a merger product appear younger by a considerable fraction of its nuclear timescale.
The strength of the rejuvenation increases for more evolved binaries, for lower mass binaries, and for binaries with a larger mass ratio. 
It is a key factor in the evolution of the binary system, and limiting the rejuvenation is expected to have large effects. 
One way to inhibit the rejuvenation is by limiting the mixing of additional fuel into the core through  convective overshooting. 
The efficiency of rejuvenation is expected to have a large impact on the evolution of a close binary. 
The size and composition of the core is related to the radius of a star, which determines the mass transfer events in the system when overflowing the Roche lobe. 

In evolutionary models of massive stars, convective boundary mixing (CBM) is still a large source of uncertainty \citep{2024ApJ...964..170J}. 
One type of CBM is overshooting.
However, both the extent and efficiency of the mixing in this region are still debated. 
In the last decade, using high-cadence photometry from space missions, asteroseismic analyses have provided accurate constraints on the convective core masses and overshooting parameters \citep{2015A&A...580A..27M, 2020MNRAS.493.4987A, 2020ApJ...904...22V, 2021A&A...650A.175M, 2021NatAs...5..715P, 2023NatAs...7..913B}. 

Convective regions in a stellar model are determined by the Schwarzschild \citep{1958ApJ...128..348S} or Ledoux \citep{1947ApJ...105..305L} criterion. 
The Schwarzschild criterion compares the adiabatic gradient with the radiative gradient to determine stability against convection. 
The Ledoux criterion takes the effect of the molecular weight gradient into account, together with the adiabatic gradient. 
A molecular weight gradient is built up in the near core region as a massive star ages and its core recedes.
During mass transfer events, the core needs to adjust, and it is possible to affect the size of the core post-mass transfer with an overshooting prescription dependent on the molecular weight gradient. 
\cite{Kai} reported promising results, showing that when modifying the CBM prescriptions in massive contact binaries, it is possible to alter their mass ratio evolution. Their work was limited to a set of models of a single initial primary mass, initial mass ratio and initial period. 
In this work, we aim to study the effect of rejuvenation on a wide parameter range of contact binary progenitors.

This paper is organized as follows. Section~\ref{sec:mixing-theory} describes the convective core overshoot and implemented limit based on the molecular gradient. 
Section~\ref{sec:methods} describes the \texttt{MESA} implementation and the models. 
Section~\ref{sec:results} contains the results. We describe one model in detail and discuss the results over the whole grid. 
Section~\ref{sec:numerical} discusses the results and investigates the numerical behaviour of the grid. 
We summarize our conclusions in Sect.~\ref{sec:conclusions}.

\section{Convective mixing}\label{sec:mixing-theory}
Convection is a mode of heat transport that occurs when material is unstable under adiabatic displacement. 
It occurs in the stellar interior when radiative heat transport becomes inefficient. 
It is a very efficient mixer that leads to chemically homogeneous regions inside stars, for example in the cores of massive stars on the main sequence. 
The size of the convective region determines the total amount of hydrogen that is available inside the core. 
This has an immediate effect on the lifetime and evolution of massive stars. 

When using the Ledoux criterion \citep{1947ApJ...105..305L},
the stability condition takes the chemical gradient $\grad_\mu \equiv \dv{\ln \mu}{\ln P}$ into account. 
A region is stable to convection if 
\begin{equation}
    \grad_{\rm{rad}}  \leq \grad_{\rm{ad}} + \frac{\phi}{\delta}\grad_{\mu},
\end{equation}
where $\phi \equiv \left( \pdv{\ln \rho}{\ln \mu}\right)_{P, T}$,  $\delta \equiv - \left( \pdv{\ln \rho}{\ln \mu} \right)_{T, \mu}$, $\grad_{\rm{ad}} \equiv \left( \dv{\ln T}{\ln P}\right)_{\rm ad}$ is the temperature gradient under adiabatic conditions, and  $\grad_{\rm{rad}} \equiv \left( \dv{\ln T}{\ln P} \right)_{\rm rad}$ is the radiative temperature gradient. 

A common prescription of convection for stellar models is local mixing length theory \citep{1953ZA.....32..135V, 1958ZA.....46..108B, bohm1976proper}. 
It calculates the efficiency of convection and the diffusion coefficient.
The efficiency of convection is a measure of the ratio of the thermal adjustment timescale to the lifetime of the convective bubble.
The diffusion coefficient quantifies the rate at which the convection makes the density, composition, and temperature gradient homogeneous. 
However, as this is a local theory, non-local effects such as  convective core overshoot must be added ad hoc.
 
As fluid parcels rise in the convective zone of a star, they gain momentum.
This non-zero momentum at the boundary keeps them from immediately dissolving, and they overshoot into the radiative zone. 
This leads to the additional mixing of envelope material into the core of the star. 
As such, the strength of the overshoot can have a large influence on the composition and evolution of the core.  

Two common ways to implement the convective overshoot are the step- and exponential overshoot. 
The former keeps the convective diffusion coefficient at the inside of the convective boundary constant into the radiative zone for some prescribed length, while the latter has the diffusion coefficient exponentially decay beyond the boundary.
\citet{2011A&A...530A.115B} calibrated the step overshoot length $(l_{P})$ as a fraction $(\alpha_{\rm{ov}})$ of the pressure scale height $(H_{\rm{P}})$:
\begin{equation}
    \label{eq:Brott}
    l_{P} = \alpha_{\rm{ov}} H_{\rm{P}}.
\end{equation}
Using data from the Survey of Massive Stars \citep{2005A&A...437..467E}, they found the best fitting result to be $\alpha_{\rm{ov}} = 0.335$.

The disadvantage of using the above formulation is that the overshooting region is entirely defined by the conditions at the convective boundary. 
It leaves no freedom for potentially changing conditions in the overshooting region.
However, since the Ledoux criterion of convection takes into account molecular weight gradients, which can inhibit convective motion, we consider the possibility for the overshoot to be similarly limited by molecular weight gradients. 

We define an upper limit $(B)$ of the molecular weight gradient, above which we do not allow the overshoot to penetrate. 
The length of the overshoot region is thus given by solving 
\begin{equation}
    \label{eq:B}
   \grad_\mu (l_{\rm{B}}) = B ,
\end{equation} where $l_{\rm{B}}$ is the gradient dependent overshoot length and $B$ is a free parameter. 
To make sure we do not extend the overshooting region, we take the overshoot length as the minimum of $l_{P}$ and $l_{\rm{B}}$:
\begin{equation}
    l_{\rm{ov}} = \min \{l_{P}, l_{\rm{B}} \}.
\end{equation}

In the case where Eq.~(\ref{eq:B}) has no solution, that is when $\grad_\mu < B$ everywhere beyond the convective boundary, we revert to using $l_P$ for the overshooting length.
We expect the evolution of our stellar models to be sensitive to the value of $B$ chosen.
A low value of $B$ will lead to a small $l_{\rm{B}}$ and as such will inhibit overshooting.  
The limit on the overshoot will inhibit the growth of the core during mass accretion, which will stall the rejuvenation of the core. 
All else being equal in a mass transfer event, the core of the accretor will be more evolved and the stellar radius will be larger.

\section{Methods}\label{sec:methods}
We computed a grid of binary evolution models using  Modules for Experiments in Stellar Astrophysics
\citep[\texttt{MESA}][]{Paxton2011, Paxton2013, Paxton2015, Paxton2018, Paxton2019, Jermyn2023} version r23.05.1 with \texttt{MesaSDK} version 23.7.3 \citep{richard_townsend_2024_10624843}.
Section~\ref{sec:micro} summarizes the microphysics, wind prescription, and binary interaction physics. Section~\ref{sec:mixing} describes the mixing parameters used, while Sect.~\ref{sec:Methods:models} gives the grid parameters and initial and termination conditions.
Finally, Sect.~\ref{sec:popsynth} details the population synthesis calculations.

\subsection{Microphysics, winds, rotation, and binary interaction}\label{sec:micro}
We used a nuclear net containing eight isotopes: $^1$H, $^3$He, $^4$He, $^{12}$C, $^{14}$N, $^{16}$O, $^{20}$Ne, and $^{24}$Mg. The appropriate nuclear reaction rates were from \citet{Cyburt2010} and \citet{Angulo1999}.
The tabulated weak reaction rates were from \citet{Fuller1985}, \citet{Oda1994}, and \citet{Langanke2000}. 
Plasma screening was included via the prescription of \citet{Chugunov2007}.
Thermal neutrino loss rates were from \citet{Itoh1996}.
The  \texttt{MESA} equation of state that we used was a blend of \citet{Rogers2002}, 
\citet{Saumon1995}, \citet{Timmes2000}, \citet{Potekhin2010}, and \citet{Jermyn2021}.
\citet{Iglesias1993, Iglesias1996} provided the radiative opacities, with low-temperature data from \citet{Ferguson2005}.
The Compton-scattering dominated regime with high temperatures was from \citet{Poutanen2017}.  
\citet{Cassisi2007} and \citet{Blouin2020} gave the electron conduction opacities.

Mass loss through winds was dependent on the surface hydrogen fraction ($X),$ following \citet{2011A&A...530A.115B}. 
The prescription of \citet{1995A&A...299..151H} was applied when $X < 0.4$.
For $X > 0.7$ and temperatures above the iron bi-stability jump \citep[as calibrated by ][]{2001A&A...369..574V}, the mass loss through winds was calculated following \citet{2001A&A...369..574V}. 
For temperatures below the iron bi-stability jump and high hydrogen surface abundance, we took the maximum of the rates described by  \citet{2001A&A...369..574V}  and \citet{1995A&A...302..811N}.
For surface abundances between these regimes, we linearly interpolated the above cases.
The wind was allowed to be accreted by the companion following the description of \citet{hurley2002evolution}.
All models assumed solar metallicity $Z_\odot = 0.0142$ and metal fractions from \citet{2009ARA&A..47..481A}. 
The rotation rate of both stars in the models was uniformly synchronized on the timescale of an orbital period, as we expected short-period binaries to be rapidly synchronized \citep{1975A&A....41..329Z}.

We implemented conservative mass transfer. 
For a semi-detached configuration, the donor was kept just below the Roche lobe and the mass was transferred onto the accretor, while in a contact phase the surface of both stars was kept on the same equipotential surface. 
We took into account the effect of tidal deformation on the internal structure of the star following the corrections of \citet{2022A&A...661A.123F}, and implemented efficient energy transfer in contact phases as in \citet{2023A&A...672A.175F}.

\subsection{Mixing parameters}\label{sec:mixing}
The convective regions were determined using the Ledoux criterion. 
We used the mixing length theory of \citet{1958ZA.....46..108B}, as described by \citet{1968pss..book.....C}, and set the mixing length parameter as $\alpha = 2$. 
We allowed for convective boundary mixing through step overshooting, where the diffusion coefficient at 0.01 pressure scale heights within the convective zone is kept constant out to $l_{\rm{ov}}$ out of the convective zone.
We limited the overshoot zone to $l_{\rm{ov}} = \min \{l_{P}, l_{\rm{B}} \}$ as described in Sect.~\ref{sec:mixing-theory}. 
We followed the model of \citet{1983A&A...126..207L} to implement semi-convection, with a high efficiency of $\alpha_{\rm{sc}}=100$ following \citet{2019A&A...625A.132S}. 
Thermohaline mixing followed the scheme of \citet{1980A&A....91..175K}, with an efficiency parameter of $\alpha_{\rm{th}} = 1$.
We included rotational mixing by including dynamical shear instability, secular shear instability, the Goldreich-Schubert-Fricke instability, and Eddington-Sweet circulation \citep{2000ApJ...528..368H}. 
The molecular gradient was calculated following \citet{Paxton2013}. 
To numerically stabilize derivatives of the composition, which may exhibit large steps over cell boundaries, a smoothing by two adjacent cells in both directions was used.

\subsection{Initialization and termination of models}\label{sec:Methods:models}

The grid consists of \num{4896} models. We considered massive binaries with an initial primary mass $M_{\rm 1, init} =  8 \text{ }M_\odot -  40 \text{ }M_\odot$ with a step $\Delta M = 2$ $M_\odot$. 
The mass ratio is the ratio of the secondary to the primary mass and takes initial values $ q_{\rm init} =\frac{M_{\rm 2, init}}{M_{\rm 1, init}} = 0.40 - 0.95$ with a step $\Delta q = 0.05$. 
We considered initial periods between $0.4$~d and $4.159$~d with a step $\Delta \log_{10} P = 0.04$. The overshoot limit defined above in Eq.~(\ref{eq:B}) was taken to be $B = 10^{-4}$. 

The models were spun up to be synchronous with the orbital period, afterwards they relaxed onto the zero-age main sequence (ZAMS), which is defined as the moment when the luminosity is within 1\% of  nuclear luminosity.
The eccentricity was kept at zero, as is expected for close binaries \citep{1975A&A....41..329Z}. 

There were multiple terminations possible for the simulations:
\begin{enumerate}
    \item $L_2$ overflow: The envelope of the contact binary overflows the $L_2$ Lagrange point, in which case the system is expected to quickly evolve to a merger.  
    \item Convergence issues: The numerical solver cannot find an acceptable solution for a finite time step.
    \item High mass transfer rate: If the mass transfer rate exceeds $10^{-2}$~$\rm{M}_\odot \rm{yr}^{-1}$, the simulation tends toward dynamical timescale mass transfer, which we do not attempt to resolve.
    \item Survival: When one of the companions leaves the main sequence, we terminate the simulation.
\end{enumerate}

If the model overflowed $L_2$ before the ZAMS condition was met or Roche-lobe overflow at ZAMS occured, we discarded it from the population. 
When any of the termination conditions was met, the simulation was quitted.
We made no distinction between the terminations and performed our further analysis on all models in the grid, regardless of termination.  

\subsection{Population synthesis}\label{sec:popsynth}
We compare the results of the performed simulation to observations of candidate massive contact binaries (see Appendix~\ref{sec:observations} for an overview). 
This was done by means of a cumulative distribution function (CDF). 

We applied the observational mass ratio $q_{\rm{obs}} = \min \left( \frac{M_2}{M_1}, \frac{M_1}{M_2}\right)$ to compare to the observational values. 
We divided the observational mass ratio and the period into bins and considered the time each model spent in each bin. 
We obtained a duration of the form $t(q_{\rm{obs}}, p_{\rm{obs}}, M_i, P_i, q_i)$, where $q_{\rm{obs}}$ denotes the observational mass ratio bin and $p_{\rm{obs}}$ denotes the period bin. 
Symbols $M_i$, $P_i$, and $q_i$ denote the initial primary mass, initial period, and initial mass ratio, respectively.
To construct a synthetic population from our grid of models, we needed to make assumptions about the distributions of initial masses, periods, and mass ratios. 

To weigh the different initial periods ($P_i$), we applied Öpik's law \citep{1924PTarO..25f...1O}. This law states that the distribution is flat in logarithmic space, which is supported by multiple surveys of massive binaries \citep{2015A&A...580A..93D, 2017A&A...598A..84A, 2021MNRAS.507.5348V, 2022A&A...658A..69B}. This means that the initial periods are weighted as 

\begin{equation}    
    \label{eq: period distribution}
    w_{P_i}(P_i)  \dd P_i = \dd \log_{10} P_i = \frac{1}{P_i \ln(10)} \dd P_i, 
\end{equation}
where $w_{P_i}$ represents the weight for the initial periods,  $P_i$. 

To correctly weigh the different initial masses $(M_i$), we applied the Salpeter initial-mass function (IMF) \citep{1955ApJ...121..161S, 2001MNRAS.322..231K, 2010ARA&A..48..339B}

\begin{equation}
    \label{eq: Salpeter}
    w_{M_i} (M_i) = M_i^{-2.35} \dd M_i,
\end{equation}
where $w_{M_i}$ represents the weight for the initial primary mass, $M_i$. 

We assumed the initial mass ratio weight to be uniform:
\begin{equation}
    \label{eq: initial mass ratio}
    w_{q_i} = q_i^0 \dd q_i = \dd q_i.
\end{equation}

As we assumed the star-formation rate (SFR) to also be uniformly distributed, we can write
\begin{eqnarray*}
    t(q_{\rm{obs}}, p_{\rm{obs}}) &\propto& \int t(q_{\rm{obs}}, p_{\rm{obs}}, M_i, P_i, q_i) w_{P_i} w_{M_i} w_{q_i} \dd P_i \dd M_i \dd q_i \\
    & = & \int t(q_{\rm{obs}}, p_{\rm{obs}}, M_i, P_i, q_i) \frac{M_i^{-2.35}}{P_i \ln(10)}\dd P_i \dd M_i \dd q_i
\end{eqnarray*}
as we integrate over the initial primary masses, periods, and mass ratios.    

The models take discrete values for the initial values as described above. 
We therefore needed to discretise the integral. 
We took each model to represent a finite area of the initial distribution. 
Let us consider the initial period distribution. 
It is uniformly distributed in logarithmic space. 
This gives
\begin{equation}
    W_{P_i} = \int_{\rm{lower}}^{\rm{upper}} \dd \log_{10} P_i = \log_{10} P_i^{\rm{upper}} -  \log_{10} P_i^{\rm{lower}}.
\end{equation}
As we have taken our initial period distribution to be evenly spread in logarithmic space, this will always take the constant value of $W_{P_i} = 0.04$. 
For the initial mass distribution, the bins at the outer ends of the model grid are chosen as such to conserve the constant bin size. 

By applying a normalization, we can consider $t(q_{\rm{obs}}, p_{\rm{obs}})$ to be probabilities $\mathcal{P}(q_{\rm{obs}}, p_{\rm{obs}})$. This allows us to compute the CDF
\begin{equation}
    \label{eq: CDF}
    \text{CDF} (q, p) = \int_{q_0 = 0.4, p_0 = 0.4 \dd}^{q, p} \mathcal{P}(q_{\rm{obs}}, p_{\rm{obs}}) \dd q_{\rm{obs}} \dd p_{\rm{obs}}.
\end{equation}The marginal distributions can be obtained by integrating over all observed periods or mass ratios. 

\section{Results}\label{sec:results}
In this section we analyse and discuss our evolution models. 
We start by discussing a single evolutionary model in detail, with a focus on the mass transfer rates and the rejuvenation. 
Finally, we compare our synthetic population against observed massive contact binaries (listed in Table~\ref{tab:pop_app:mw_systems}).

\subsection{Detailed model evolution}\label{sec:results:model}
Here, we describe in detail an example of our case A interacting models. 
We selected the initial conditions of $M_{1, \rm{initial}} = 24 \text{ }M_\odot$, $P_{\rm{initial}} = 1.377$~d, and $q_{\rm{initial}} = 0.8$. 
As for all the models, we set the overshoot limit at $B = 10^{-4}$. 
To set a baseline model, we selected the \texttt{MESA} models of \citet{PhDFabry} and \citet{Fabry2024}. 
They are set up in a similar way, with the exception that they follow the overshooting  $l_{P} = 0.335 H_{P}$ as defined in Eq.~(\ref{eq:Brott}). 
The evolution of these models closely follows those of \citet{2021MNRAS.507.5013M}. 
We refer to this model as the Full Overshoot (FO) model. 

\begin{figure}
    \centering
    \includegraphics[width=\linewidth]{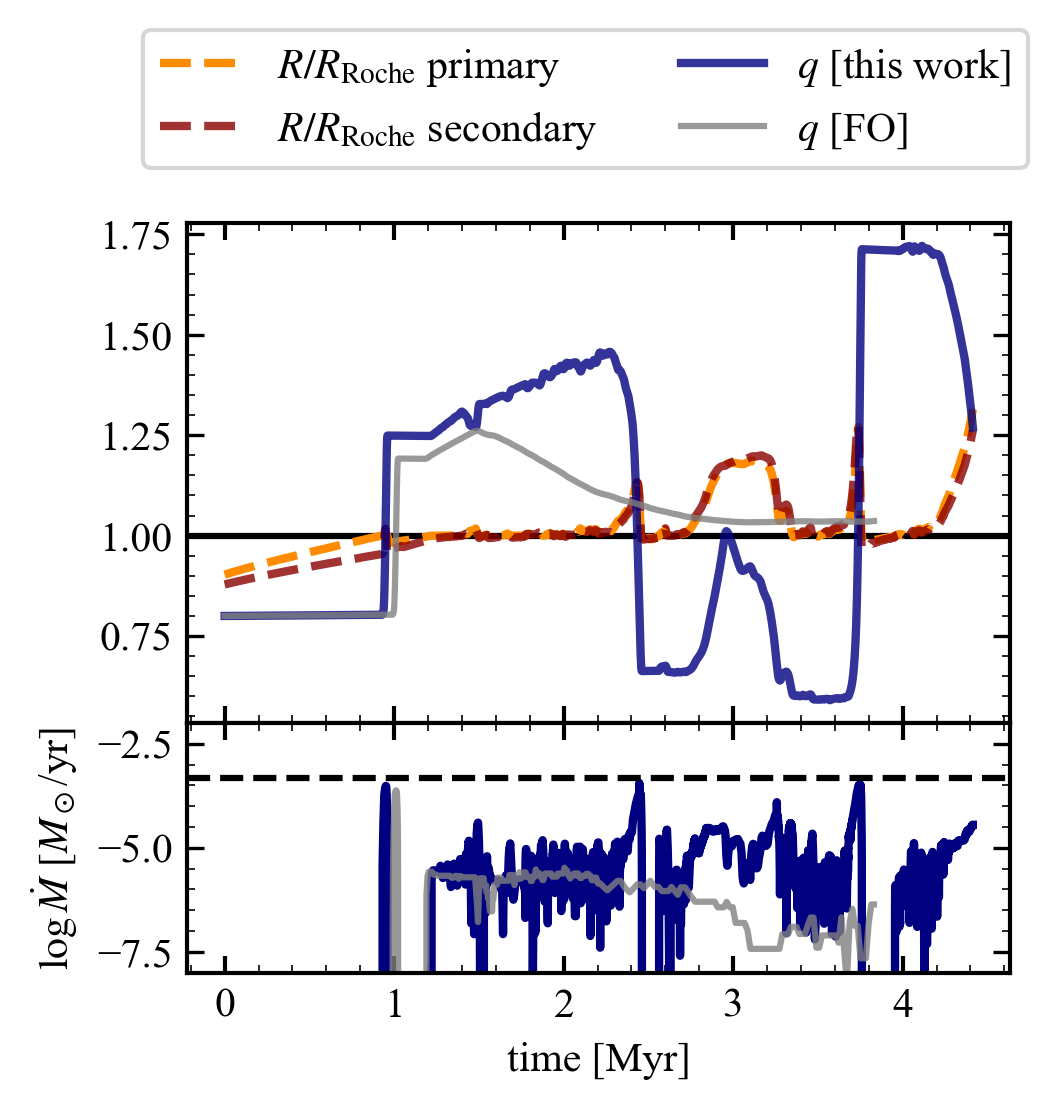}
    \caption{Example of the mass ratio evolution with initial conditions of $M_{1, \rm{initial}} = 24 \text{ }M_\odot$, $P_{\rm{initial}} = 1.377$ d, and $q_{\rm{initial}} = 0.8$. 
    Top panel: Mass ratio evolution of the system. %} 
    The coloured lines show the new model, the grey line shows the FO mass ratio evolution. The dashed lines show the stellar radii normalized by the respective Roche lobe radii.
    Bottom panel: Mass transfer rate of the system of the new model (blue) and the FO model (grey). The dashed black  line denotes the initial thermal mass transfer rate of the secondary.} 
    \label{fig:model-evolution}
\end{figure}

Figure~\ref{fig:model-evolution} shows the mass ratio evolution of the model of this work together with the mass ratio evolution of the FO model. 
The FO model closely follows System 2 of \citet{2021MNRAS.507.5013M} as expected, with a brief contact phase (fast case A mass transfer) around 1 Myr followed by a short detached phase. 
Afterwards, a semi-detached phase follows as mass is transferred from the primary to the secondary (slow case A). 
A new contact phase is formed when the secondary reaches its Roche lobe  and the system equalizes over a nuclear timescale. 
The stars merge around 3.9~Myr after the ZAMS. 

The new overshoot limited model deviates considerably from the FO model. 
This is immediately noticeable on Fig.~\ref{fig:model-evolution}. 
We  go over some important differences. 

The new model has a longer lifetime by about 0.6 Myr. 
Additionally, the fast case A mass transfer starts about 40 kyr earlier than the reference model. 
As we limit the overshoot, we limit the mixing at the core boundary. 
As less envelope hydrogen is mixed into the core, the stars will age faster. 
The faster ageing is accompanied by expansion, and the donor reaches the Roche lobe faster. 

The fast case A leads to a larger mass ratio before detachment. 
While the FO model has $q = 1.19$ after the fast case A, the new model is at mass ratio $q=1.25$. 
The slow case A that follows pushes the mass ratio further from unity and takes about 1.1 Myr. 
This slow case A is notably not smooth. 
This seems to be a consequence of the system repeatedly going in and out of a contact phase. 
We suspect this behaviour to be of numerical origin.
The numerical behaviour of our models is discussed in Sect.~\ref{sec:numerical}.

After the slow case A mass transfer event, multiple rapid mass ratio inversions take place over multiple contact phases and the mass ratio is pushed to more extreme values. 
A final contact phase brings the mass ratio down, but due to $L_2$ overflow, the system is considered to be merged. 

The new model has a similar early evolution to the FO models, but deviates strongly from about 1.5 Myr onwards. 
The slow convergence towards unity is replaced by multiple rapid mass transfer events.
Each transfer seems to deviate the mass ratio further from unity. 
The evolution of the system in the HR diagram is shown in Appendix~\ref{sec:app:HR}.

We see that for the fast case A and following slow case A mass transfer events up to around 1.5~Myr, 
both models seem to show similar mass transfer rates. 
This is confirmed by the bottom panel of Fig.~\ref{fig:model-evolution}, which shows the mass transfer rate for the new model in blue and the mass transfer for the FO model in grey.
Similar to above, the plot shows staggered behaviour as a consequence of the numerical implementation. 
We do see, however, that overall the mass transfer rate is higher in the new model especially at later times in the evolution, which we suspect to 
be a consequence of the larger radii that accompany the more evolved cores. 
The three mass ratio inversion events are accompanied by strong mass transfer rates, which approach the initial thermal mass transfer rate of the secondary $\log \Dot{M}_{\rm KH, init, 2} = \frac{M_{\rm init, 2}}{\tau_{\rm KH, 2}}$ (black line), where $\tau_{\rm KH}$ is the Kelvin-Helmholtz timescale.

\subsection{Rejuvenation}
\begin{figure}
    \centering
    \includegraphics[width=\linewidth]{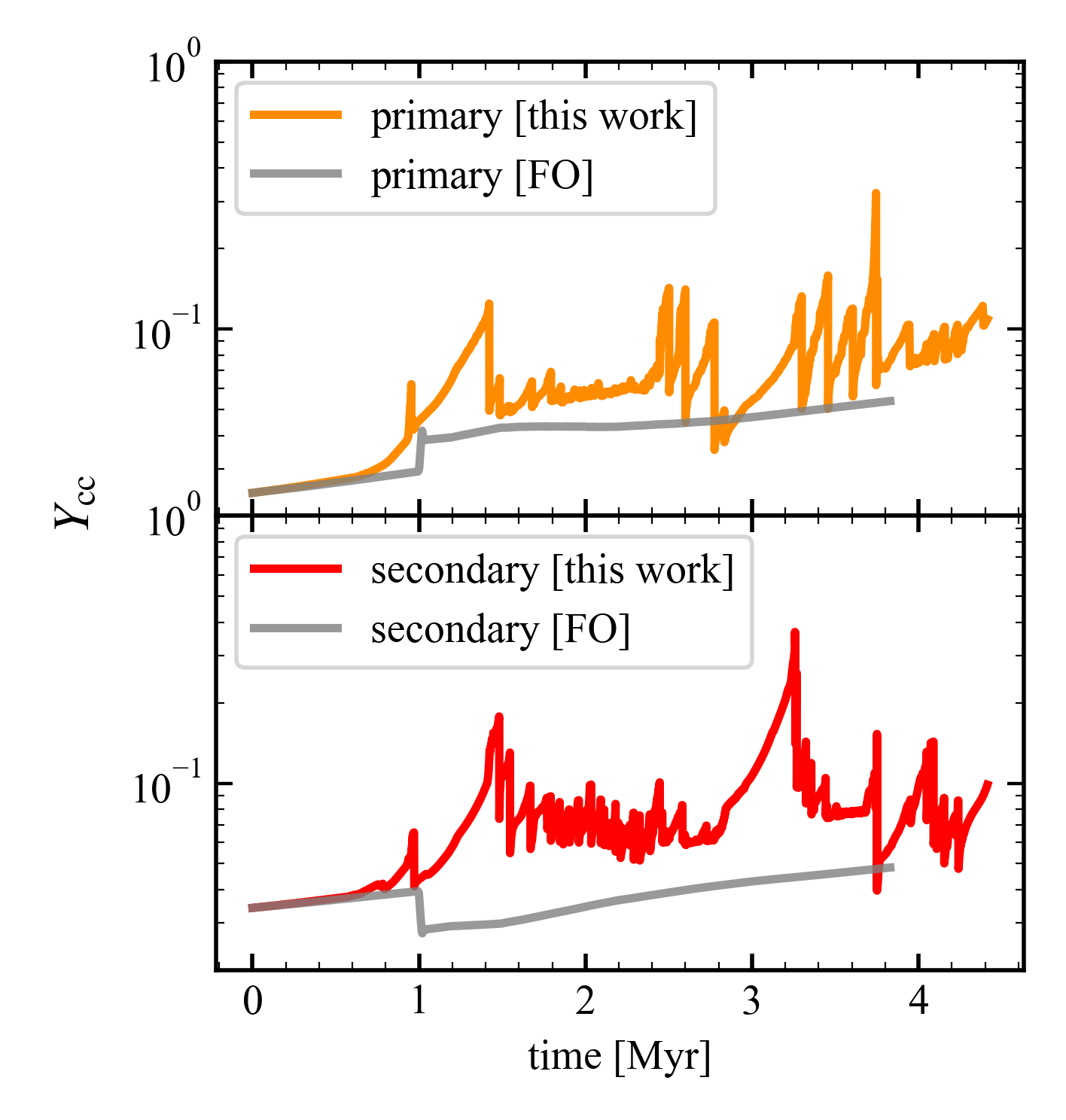}
    \caption{Evolution of the helium mass fraction in the core of the star. The upper plot shows the primary star, the lower plot shows the secondary star. The grey line denotes the FO model. }
    \label{fig:rejuvination}
\end{figure}

The new method inhibits convective boundary mixing when the molecular gradient becomes larger than our boundary value. 
Less near-core mixing will lead to a smaller rejuvenation of the core. 
This can be seen in Fig.~\ref{fig:rejuvination}, which shows the evolution of the helium mass fraction of the convective cores ($Y_{\rm{cc}}$) of both the primary and the secondary (for the same model presented in Fig.~\ref{fig:model-evolution}). 

The effect of our model is clear before the first contact phase is initiated at 1~Myr, as the helium mass fraction rises more rapidly than in the FO case. 
This behaviour continues after the first fast mass transfer event. 
The mass transfer event seems to lower the molecular gradient near the core as a rejuvenation happens for both companions. 
Similar behaviour is present throughout the lifetime of the system, where the helium core mass fraction grows steadily until the convective core overshoot can break through the molecular gradient. 
This then leads to a quick rejuvenation of the core. 
The staggered behaviour is an effect of the current implementation. 
We make use of a step overshoot with a hard limit on the molecular gradient (see Sect. \ref{sec:numerical} for a discussion of the numerical behaviour).  

In general, both helium mass fractions are larger than in the FO models, indicating that our overshooting scheme reduces the efficiency of rejuvenation, which has a direct effect on the evolution. 
As discussed in the previous section, the slow case A mass transfer is extended and multiple inversions of the mass ratio take place. 
 
\subsection{Full grid results}
\begin{figure}
    \centering
    \includegraphics[width=\linewidth]{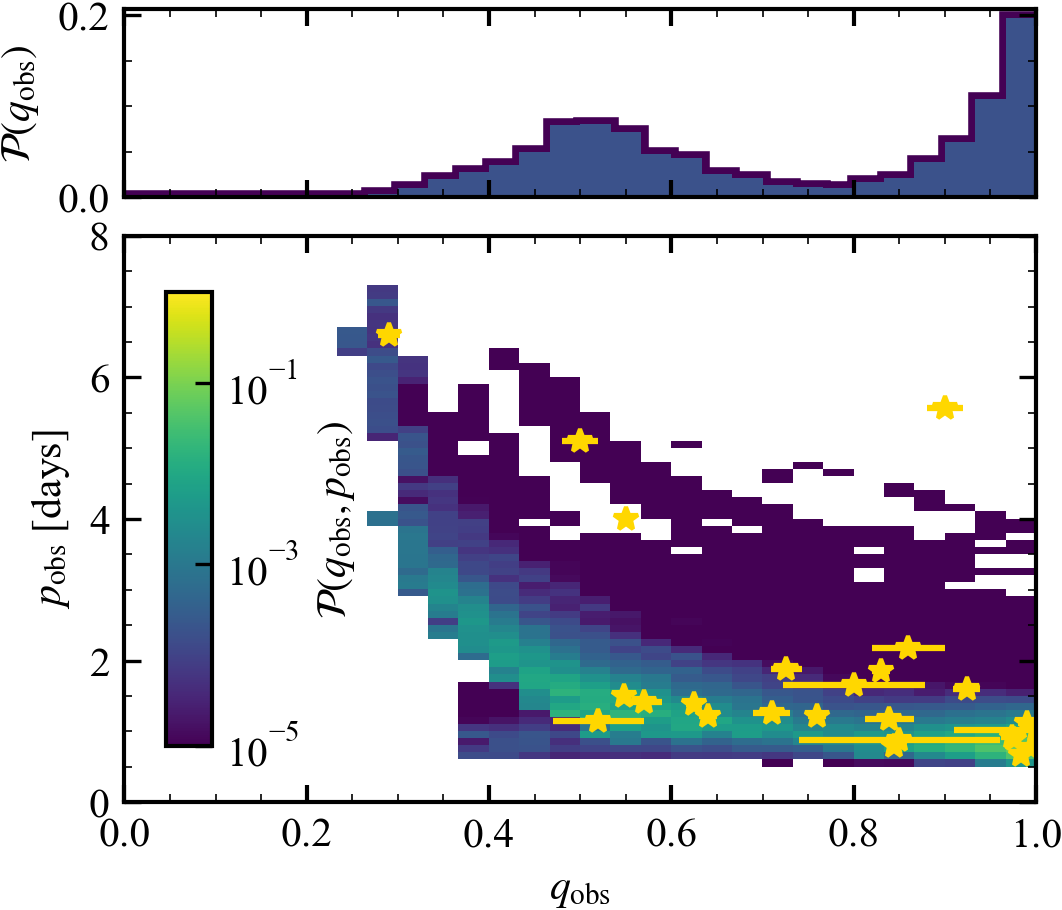}
    \caption{Two-dimensional distribution of $ t(q_{\rm{obs}}, p_{\rm{obs}})$. The gold stars denote the observed massive contact binaries. The upper plot shows the PDF of $ \mathcal{P}(q_{\rm{obs}})$.}
    \label{fig:2Dhist}
\end{figure}

\begin{figure}
    \centering
    \includegraphics[width=\linewidth]{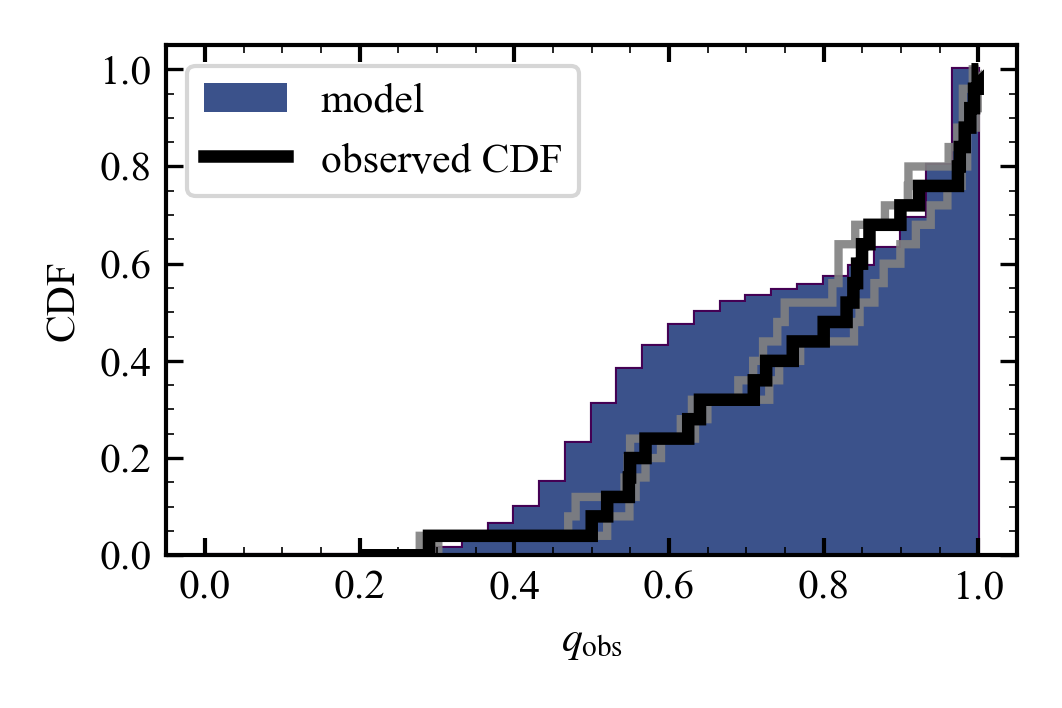}
    \caption{Cumulative distribution function of the observational mass ratios. The black line denotes the CDF of the observed contact binaries. 
    The grey lines denote CDFs of the observations with the mass ratios shifted by their errors.  }
    \label{fig:qdist}
\end{figure}

\begin{figure}
    \centering
    \includegraphics[width=\linewidth]{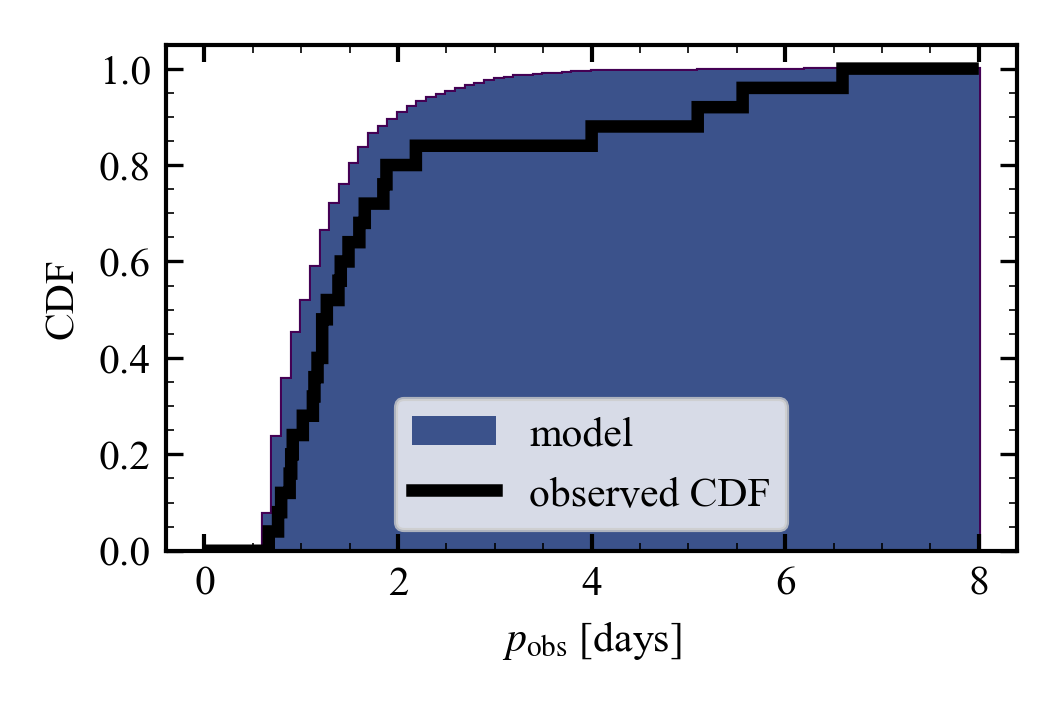}
    \caption{Cumulative distribution function of the observational periods. The black line denotes the CDF of the observed contact binaries.}
    \label{fig:pdist}
\end{figure}

We evolved \num{4896} models (for grid description, see Sect. \ref{sec:Methods:models}) and analysed the contact phases.  
Of the \num{4896} models, about 27.7\% ran into numerical trouble. 
Most often, the time step became too small because \texttt{MESA} could not find a convergent model.
This means that 72.3\% of the models evolved as expected. 
Of all the models, 14.4\% models contained at least one star that left the main sequence before merging. 
These are mostly models with large initial periods. 
The models with $L_2$ overflow before the ZAMS account for 21.7\%. 
These all have initial periods below one day. 
Only 5.4\% of the models attained the high mass transfer termination as described above. 
In total, 1507 models reached $L_2$ overflow on the main sequence (30.8\%). 

The weighted distribution $ \mathcal{P}(q_{\rm{obs}}, p_{\rm{obs}})$ is visualized in Fig.~\ref{fig:2Dhist}. 
This plot shows the probability distribution of the time spent in each observed mass ratio and period bin. 
The known contact binaries are denoted by gold stars. The uncertainties on the periods are  negligible. 
The probability distribution for the mass ratio is shown in the upper plot. 
This is obtained by integrating over all period bins. 

\begin{figure*}
    \centering
    \includegraphics[width=\textwidth]{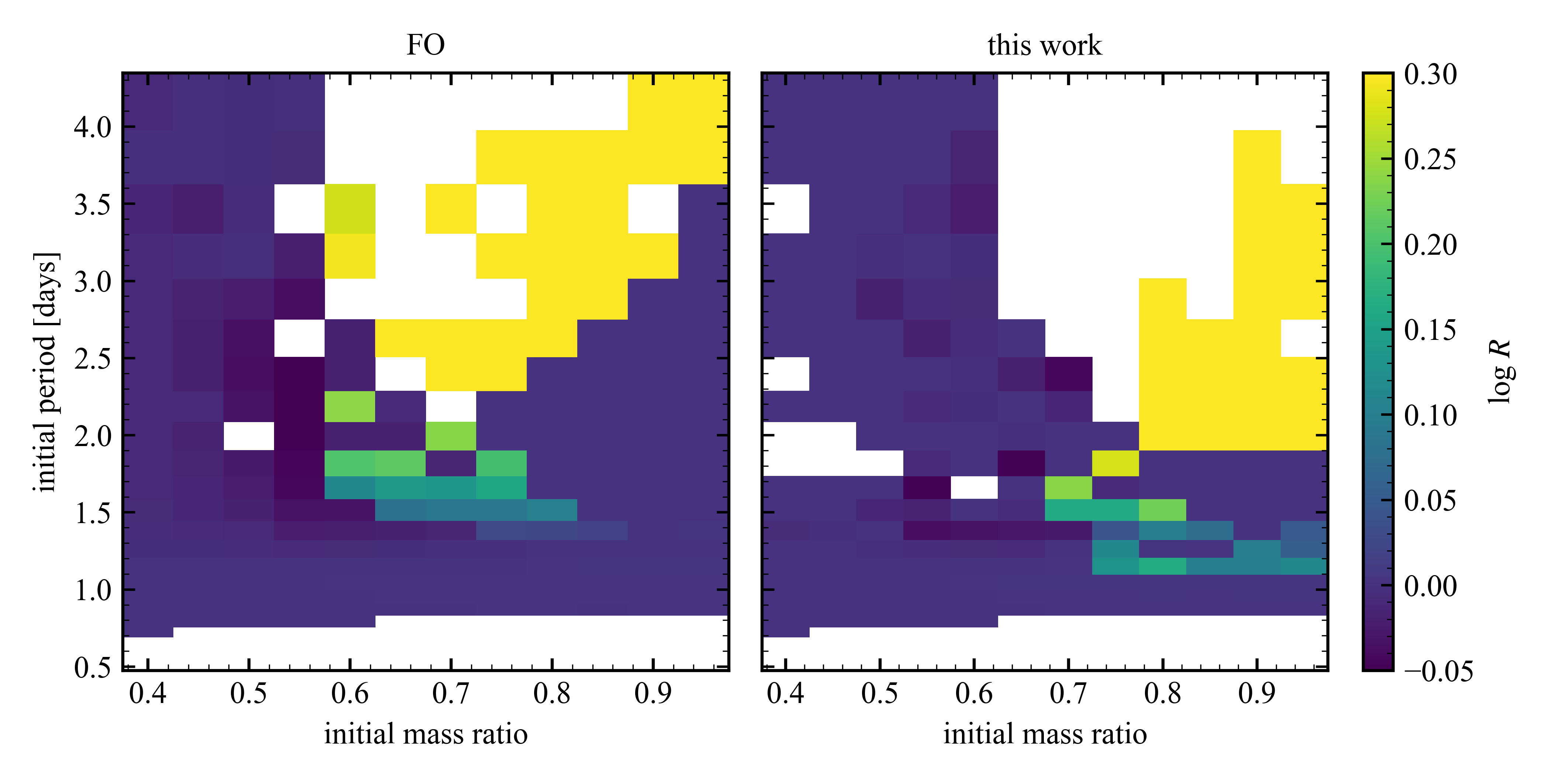}
      \caption{Ratio of the helium mass fraction of the core at the moment the initial secondary becomes the donor, and before the first interaction for the grid models with $M_{\rm 1, init} = 24$~$M_\odot$. The left plot gives the values for the Full Overshoot (FO) models of \citet{Fabry2024} and the right plot gives the parameter values for the models of this work.}
        \label{fig:rejuvenation-fullgrid}
\end{figure*}

The 2D histogram in Fig.~\ref{fig:2Dhist} has the same shape as that seen in \citet{2021MNRAS.507.5013M} and \citet{PhDFabry}, as apparent from the strong correlation between $q$ and $p$. 
However, the mass ratio distribution in this work reaches more extreme values.
The probability distribution of $q_{\rm{obs}}$ is less skewed to unity compared to previous studies. 
We have a clear bimodal distribution with a secondary peak around $q_{\rm{obs}} = 0.5$. 
The bimodality is also clearly visible in Fig.~\ref{fig:qdist}, which shows the cumulative distribution of the observational mass ratios. 
The CDF of the observations is shown in black. 
Figure~\ref{fig:pdist} shows the cumulative distribution function of the observed period. 
The general trend of the periods is explained by our model, although we overestimate the short-period systems. 
We suspect this is a mass effect, as the observations contain more O+O systems, while the population is dominated by B+B systems on the grounds of the initial mass function.

Figure~\ref{fig:2Dhist} shows that quite a few observations fall in regions of low probability. 
This indicates that the model still does not fully explain the observations. 
A Kolmogorov-Smirnov test to compare the observed sample to the predicted distribution can be performed on both the marginal period and the mass ratio distribution. 
For the period distribution, we obtain a $p$-value of 0.024. 
For the mass ratio distribution, we obtain a $p$-value of 0.058. 
Considering a confidence level of 95\%, we can reject the period distribution, but not the mass ratio distribution. 
However, this is a large increase from a similar test performed by \citet{PhDFabry} and \citet{Fabry2024} who report a $p$-value of $6.2 \times 10 ^{-7}$. 

To quantify the different rejuvenation efficiency between the FO models and this work, we consider the ratio of the helium mass fraction of the core 
at the moment the initial secondary becomes the mass donor of the system\footnote{We ignore small numerical jumps in the evolution.}, and before the first interaction:
\begin{equation}
    R = \frac{Y_{\text{core, initial secondary becomes donor}}}{ Y_{\text{core, before interaction}}}.
\end{equation}
We calculate the ratio for a slice of the grid above with a single initial mass value $M_{\rm 1, init} = 24$~$M_\odot$ and display the resulting values in Fig.~\ref{fig:rejuvenation-fullgrid}. 
For the models with an initial mass ratio close to unity and a short initial period, we expect the ratio $R$ to be sensitive to the core boundary processes. 
We see that in this regime, the new models have noticeably higher ratios, indicating a less efficient rejuvenation. 
As the inhibited overshoot leads to smaller convective cores, we expect there to be lower mixing of the envelope hydrogen into the core, which is especially relevant for the accretor during mass transfer of the system and leads to weaker rejuvenation.
We suspect this to be the dominant effect for short initial period systems on the ratio $R$. 
For longer initial periods, a higher helium core mass fraction and smaller core radii are expected when the mass transfer occurs, as the stars are further along their evolution. 
This leads to a higher denominator in the ratio, which leads to lower values.

\subsection{Surviving systems}
\begin{figure}
    \centering
    \includegraphics[width=\linewidth]{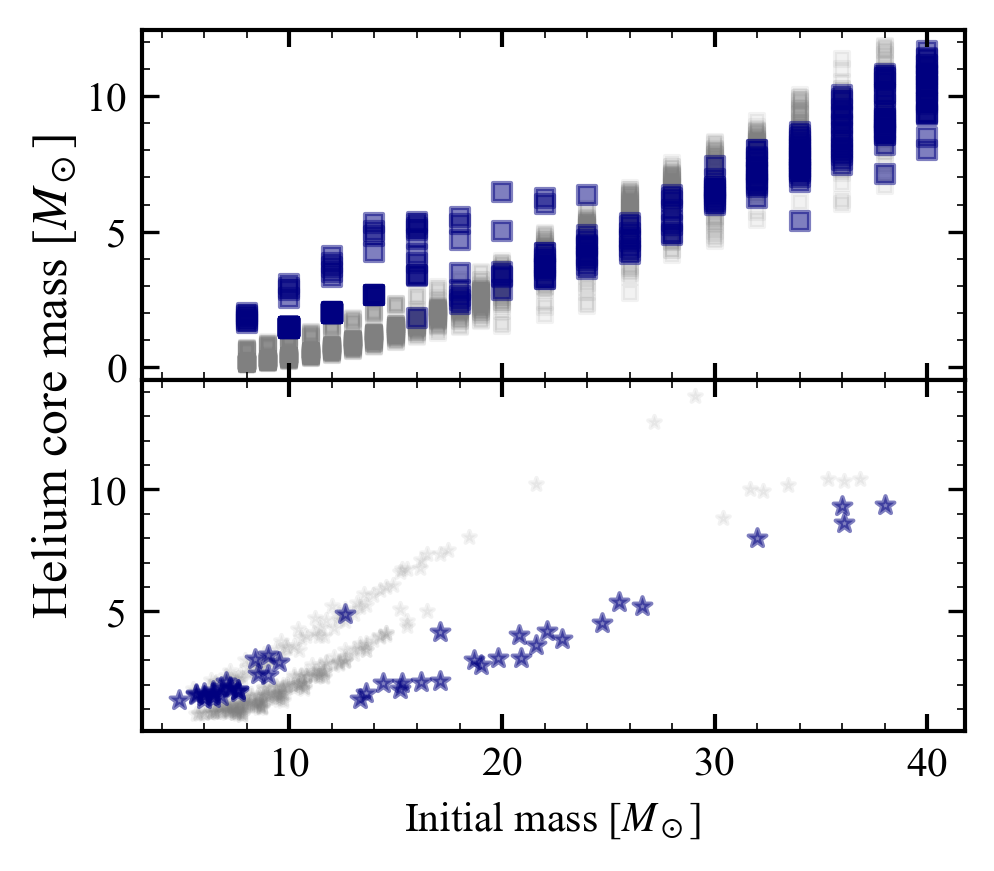}
    \caption{Helium core masses of the surviving systems compared to the initial mass of the stars. The new models are shown in blue and the FO in grey. A square (top panel) is used when the primary leaves the main sequence, a star (bottom panel) when the secondary leaves the main sequence. }
    \label{fig:hecore}
\end{figure}

Of the full grid, 14.4\% of the models were terminated due to one companion leaving the main sequence. 
These surviving systems are of interest as they contain gravitational-wave progenitors, and the remaining helium core masses are of special interest.
We compare the helium core masses of the surviving systems to the helium core masses of the FO models, using the models of the grid presented in \citet{Fabry2024} within the initial condition ranges of this study.
The helium core masses of both models are shown in Fig.~\ref{fig:hecore}, with the new models in blue and the FO models in grey compared to the initial mass. 
We show the core masses for the companion leaving the main sequence. 
For the FO models, we took the helium core to be the convective core mass at termination, while for the new models, we defined the helium core masses to be the mass coordinate where the hydrogen mass fraction drops below 10\%. 

For both implementations, the primary stars are most often the first to leave the main sequence, with the overshoot limited implementation generally leading to similar helium core masses, especially at the higher mass. 
At the lower initial mass end, we see a difference in the helium core masses, which is most likely due to the different definition used for the helium core mass and, as such, we see no significant difference in the helium core masses of the initial primaries that survive the main sequence.
The bottom panel suggests a systematic decrease in the helium cores of the secondaries that leave the main sequence, which is expected due to inhibited core rejuvenation as the accretor during case A mass transfer. 

\section{Discussion and numerical behaviour}\label{sec:numerical}
The above results on the predicted mass ratio distribution of massive contact binaries are different to what is currently presented in the literature. 
The additional dependence of the overshoot on the chemical gradient limits the rejuvenation of the accretor and affects the size of the receding convective cores. 
The presented method overpredicts systems between the observed mass ratios $q_{\rm{obs}}=0.4-0.6,$ with a clear peak in the probability distribution function at $q_{\rm{obs}}=0.5$. 
The value of the overshoot limit $(B)$ chosen is expected to have a large effect on the behaviour of the system; the exact impact of this value should be studied further. 

One large caveat is the precise implementation of the dependence on the molecular gradient. 
As this is obtained through numerical differentiation, there could be a large dependence on the refinement of the grid. 
This is clear from the following parameter study. 
To study the effects of varying the smoothing parameters, we selected the model with initial values $ M_{\rm 1, init}
 = 24 \textrm{ }M_\odot$, $P_{\rm{initial}} = 1.377$ d, and $q_{\rm{initial}} = 0.8$, the same initial values as the model shown in Sect. \ref{sec:results:model}.

The parameter with the largest effect on the numerical behaviour of the MESA models is the smoothing used to calculate composition gradients (called \texttt{num\_cells\_for\_smooth\_gradL\_composition\_term}\footnote{\url{https://docs.mesastar.org/en/23.05.1/reference/controls.html\#num-cells-for-smooth-gradl-composition-term}}), which takes a value of two in the grid presented above. 
This parameter defines the number of grid cells taken on each side over which a weighted average is taken to compute $\nabla_\mu$. 

\begin{figure}
    \centering
    \includegraphics[width=\linewidth]{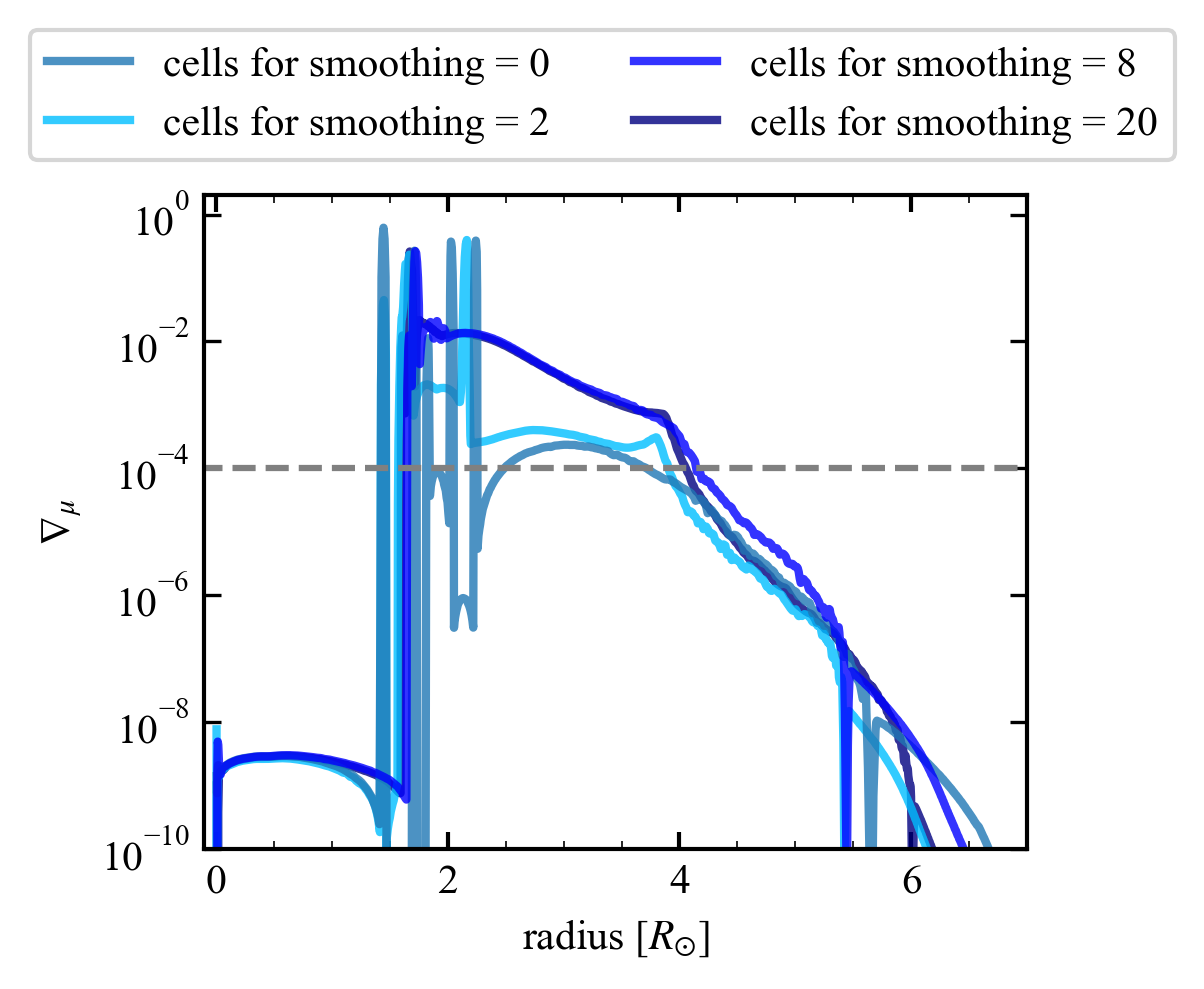}
    \caption{Molecular gradients of the primary star for different values of the smoothing parameter. The model has initial values $M_{1, \rm{initial}} = 24 \textrm{ }M_\odot$, $P_{\rm{initial}} = 1.377$~d, and $q_{\rm{initial}} = 0.8$. The profiles are selected close to 1.75~Myr. The dashed grey line denotes the overshoot limit $B = 10^{-4}$. }
    \label{fig:B-comparison}
\end{figure}

Figure~\ref{fig:B-comparison} shows the molecular gradient of the primary star at 1.75 Myr after ZAMS, which puts the model in the slow case A mass transfer event. 
The figure shows that there are large differences depending on the smoothing parameter. 
As expected, a low number of cells leads to peaked molecular gradients with smaller cores, while a larger number of cells leads to a smoother molecular gradient. 
However, it is unclear  what the optimal value for this parameter should be. 
The currently implemented default of \texttt{MESA} is 3, which is likely to increase the numerical stability of the evolution models.

\begin{figure}
    \centering
    \includegraphics[width=\linewidth]{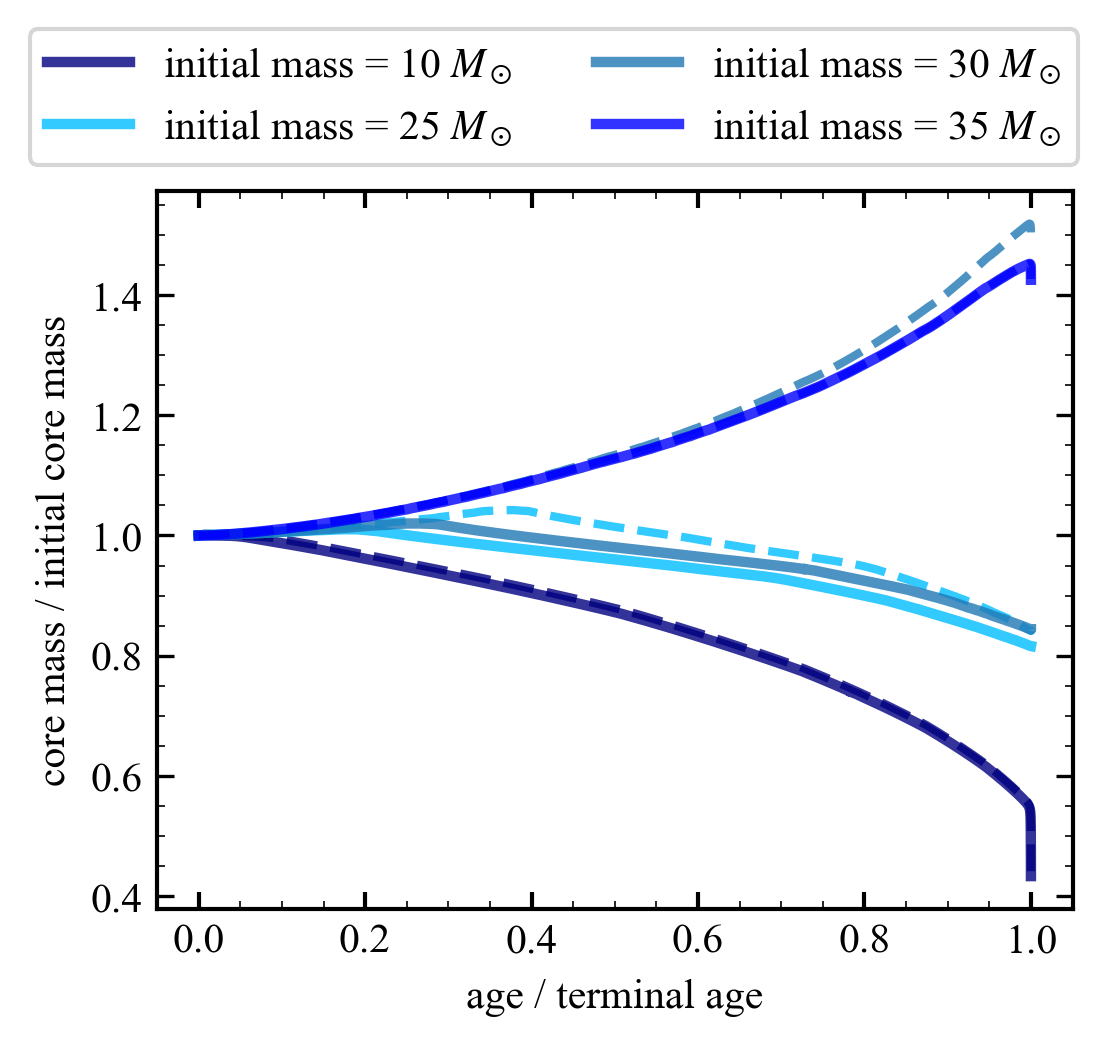}
    \caption{Core mass evolution of a single star model with fixed rotational velocity for different initial masses. The solid lines denote \texttt{num\_cells\_for\_smooth\_gradL\_composition\_term}~=~2, and the dashed lines have \texttt{num\_cells\_for\_smooth\_gradL \_composition\_term}~=~5. }
    \label{fig:mass-resolution}
\end{figure}

\begin{figure}
    \centering
    \includegraphics[width=\linewidth]{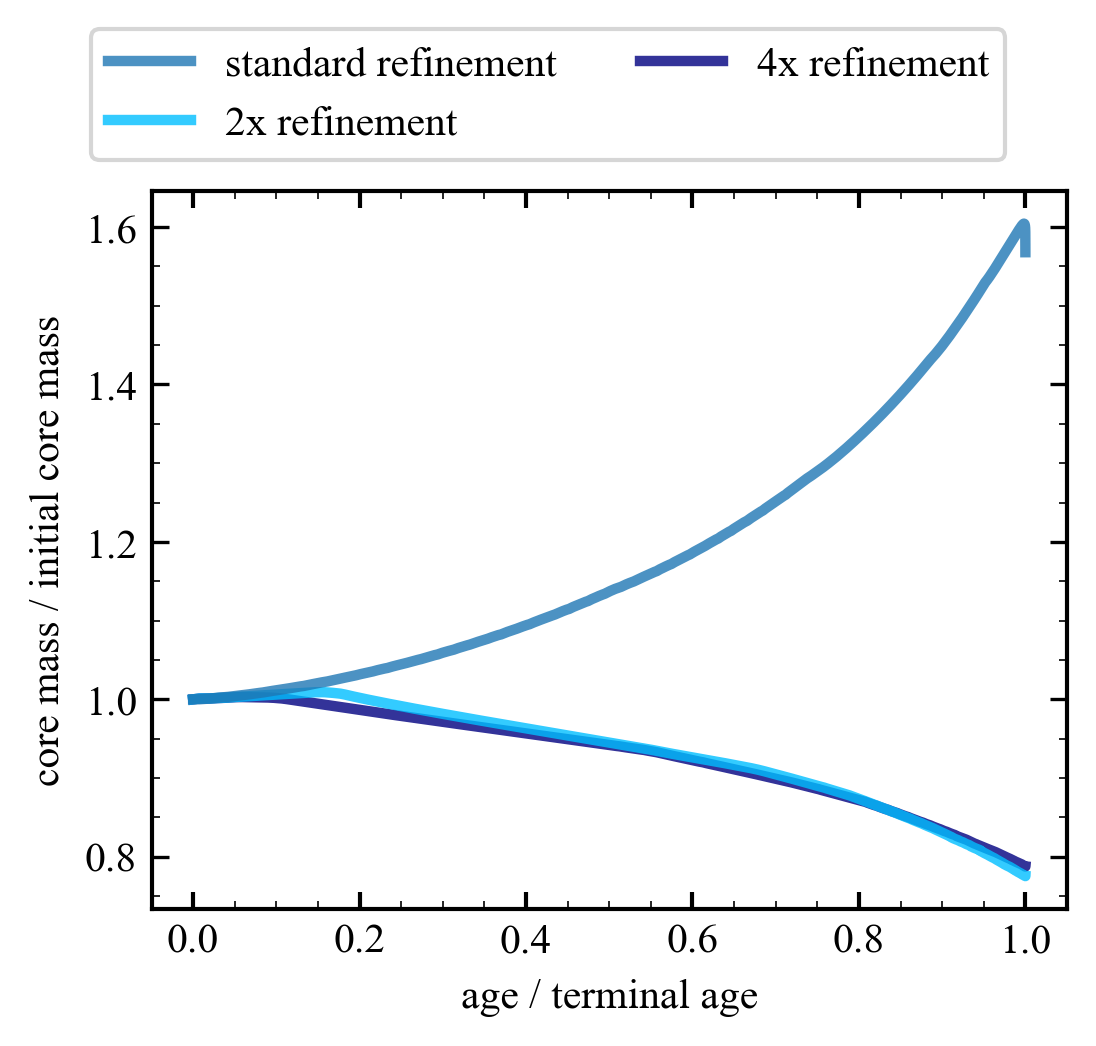}
    \caption{Core mass evolution of a single star model with different mesh refinements with an initial mass of $M = 24 \textrm{ }M_\odot$.  All models apply \texttt{num\_cells\_for\_smooth\_gradL\_composition\_term} = 10.}
    \label{fig:mesh-resolution}
\end{figure}

To investigate the numerical behaviour of the smoothing further, single star models of masses 10~$M_\odot$, 25~$M_\odot$, 30~$M_\odot$, and 35~$M_\odot$ were computed with a fixed rotation rate of $0.75 \omega_{\rm{crit}}$ and with the microphysics given in Sect.~\ref{sec:micro}.
In these models, we applied no overshoot limiting to isolate as best as possible the effect of the smoothing parameter. 
This was done to approach the situation of a binary model without considering any mass transfer. 
Figure~\ref{fig:mass-resolution} gives the core mass evolution of the different single star models. 
The solid lines give \texttt{num\_cells\_for\_smooth\_gradL\_composition\_term}~=~2 and the dashed lines give \texttt{num\_cells\_for\_smooth
\_gradL\_composition\_term}~=~5.

The core mass shrinks for the models with an initial mass $M_{\rm{initial}} \leq 30$~$M_\odot$ and a value of 2 for the smoothing parameter.
The model with an initial mass $M_{\rm{initial}} = 35 \text{ }M_\odot$ has an increasing core mass for the same model parameters. 
The initial masses for which such an increase happens are dependent on the value of the smoothing parameter. 
This can be seen by the model with an initial mass of $30 \text{ }M_\odot$ that decreases for \texttt{num\_cells\_for\_smooth\_gradL\_composition\_term}~=~2, but increases when the value is set to 5. 

The large increase in the core mass during the main sequence is a numerical effect. 
This can be seen in Fig.~\ref{fig:mesh-resolution}, which shows different mesh refinements for the single star model with an initial mass $M_{\rm{initial}} = 24 \text{ }M_\odot$ and \texttt{num\_cells\_for\_smooth\_gradL\_composition\_term}~=~10. 
For a larger refinement,  core growth does not happen, and the evolution reverts to the core receding, as expected for massive stars. 

We find that the high mass models are sensitive to the smoothing on the molecular gradient. 
The sensitivity increases with higher mass, where the core can artificially increase in mass.
This can be helped with a larger mesh refinement. 
These findings are similar to those of \citet{2016ApJ...817...54M}  for the lower mass stars with a convective core (1.2 $M_\odot$ - 1.7 $M_\odot$.) 
The authors find that the core artificially shrinks for a model with smoothing enabled, which is unexpected for stars in that mass range. 
In the mass regime studied here, the cores artificially grow.
In both cases, the cores do not behave correctly when implementing a non-zero value for the smoothing parameter. 
We note that, similar to \citet{2016ApJ...817...54M}, we considered the Ledoux criterion to determine the core size. 

The grid implemented \texttt{num\_cells\_for\_smooth \_gradL\_composition\_term} = 2.
From this discussion, we see that models with an initial mass over 30~$M_\odot$ can behave unexpectedly. 
However, excluding these models has little impact on the probability distribution of the observed mass ratio and periods, since the IMF heavily favours lower mass stars. 
We therefore chose to include all models in the grid in the analysis.

\section{Conclusions}\label{sec:conclusions}
In this work, we investigated the effects of limiting the convective core overshoot on a massive contact binary population.
We limited the overshoot based on the molecular weight gradient.
This was accomplished by using a step overshoot with an additional requirement that the molecular weight gradient not exceed the threshold value $B = 10^{-4}$. 
The goal of this implementation was to limit the rejuvenation efficiency during mass transfer.

We calculated a grid of \num{4896} models with different initial masses, mass ratios, and periods. 
This allowed us to investigate the resulting population and compare with observations. 
Our main conclusions are as follows.
\begin{enumerate}
    \item By limiting the overshoot, the mixing of envelope material into the core is reduced. 
    This inhibits the rejuvenation present in contact binaries. 
    We notice a lower rejuvenation that is staggered due to the hard limit presented by the current implementation. 
    Throughout the grid, the rejuvenation is less efficient. 
    \item Because of the reduced rejuvenation, we considerably change the evolution of a massive contact binary, with multiple contact phases and more extreme mass ratios. 
    This is in contrast to the previously presented models in the literature. 
    The population synthesis shows an additional peak in the PDF of the mass ratios, and we overpredict the number of unequal-mass contact binaries.  
    \item High mass stellar models are sensitive to smoothing on the Ledoux gradient.
    This sensitivity increases with stellar mass and can make the core artificially grow. 
    When using smoothing to aid numerical stability, the mesh density of the stellar models needs to be high enough to assure the correct behaviour of the convective boundary.
\end{enumerate}
We showed that the implementation of stellar physics can have large impacts on binary population synthesis. 
Follow-up work could include extending the grid to different values of the overshoot limit, $B$. Additionally, a more gradual dependence on the molecular weight gradient could be explored.
Other processes to inhibit the core rejuvenation of the accretor, such as magnetism and radiative transfer, could also be considered. 

Nevertheless, to make progress in the understanding of rejuvenation, and thus of the evolution of massive contact binaries, we lack observationally motivated prescriptions for its efficiency. 
Asteroseismology has mainly focused on single (rotating) stars to obtain reliable convective core masses, but as of yet no asteroseismic properties of accretors in binary systems have been published. 
Such efforts will likely bear fruit, however, as \citet{2024A&A...682A.169H} have identified that merged stars carry asteroseismic imprints that are distinguishable from true single stars.

Finally, we note that when computing massive star models, the mesh refinement and Ledoux smoothing should be carefully considered. 
This is of high importance for the asteroseismic estimation of the core overshoot and the core mass of massive stars \citep[e.g. ][]{2003Sci...300.1926A, 2004MNRAS.355..352A, 2006ApJ...642..470A, 2006A&A...459..589M, 2023NatAs...7..913B, 2024arXiv240806097F}. 
In particular, $\beta$ Cephei pulsators are in the regime where this numerical behaviour is present.
As shown, the choice of smoothing parameter value can strongly affect the modelled core size. 
Therefore, when comparing models to asteroseismic observables, systematic errors could be introduced by this artificial smoothing.

\begin{acknowledgements}
M.F. thanks the Flemish research foundation
(FWO, Fonds voor Wetenschappelijk Onderzoek) PhD fellowship
No. 11H2421N for its support.
The data was processed and visualised using \texttt{Matplotlib} \citep{Hunter:2007}, \texttt{SciPy} \citep{2020SciPy-NMeth} and \texttt{NumPy} \citep{harris2020array}.
\end{acknowledgements}

\bibliographystyle{aa} % style aa.bst
\bibliography{ads} % your references Yourfile.bib

\appendix
\section{HR diagram of the example evolution}\label{sec:app:HR}
\begin{figure}[h!]
    \centering
    \includegraphics[width=\linewidth]{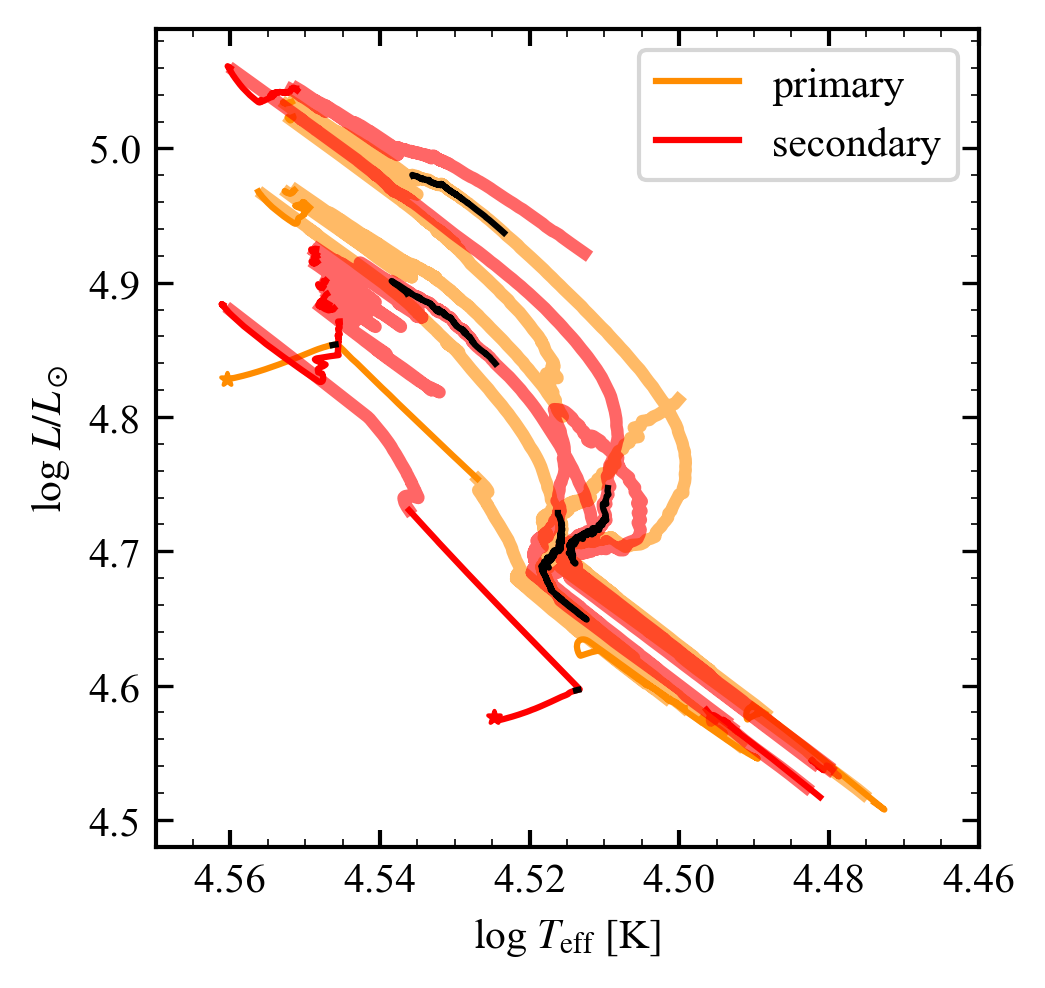}
    \caption{HR diagram of the model presented in Sect.~\ref{sec:results:model} for both the primary (orange) and the secondary (red). The contact phases are shown with a lower opacity. The ZAMS is highlighted by a star. The black lines denote the three mass transfer events approximating thermal mass transfer rate.}
    \label{fig:HRD}
\end{figure}

The evolution of a contact binary system is heavily influenced by the adjusted rejuvenation efficiency presented above. 
Figure~\ref{fig:HRD} shows the HR diagram of the model presented in Sect.~\ref{sec:results:model}, where we see the evolution of the primary (orange) and the secondary (red).
The tracks in the HR diagram are strongly deviated from models previously presented in the literature due to the multiple strong mass transfer events.
We recognize the initial discontiuity of the case-A mass transfer, where the first contact phase takes place.
The inversions of the mass ratios are clearly visible in the global evolution of the system, with the primary and secondary track switching between the upper left and lower right of the HR diagram. 
Around 3~Myr, the system undergoes a `bump' in the mass ratio, where there is a quick evolution to equal mass ratio, but it does not invert and decreases back to a lower mass ratio.
This characteristic is also visible in the HR diagram, where there are two distict loops halfway through the evolution, which cross around $\log T_{\rm eff}\sim 4.515$ and $\log L/L_\odot \sim 4.8$.

\onecolumn
\section{Observations}\label{sec:observations}

\begin{table*}[h!]\small
    \centering
 \caption{Parameters of observed, massive (near-)contact systems in the Milky Way, M31, and the Magellanic Clouds.}
                \label{tab:pop_app:mw_systems}
                \begin{tabular}{l| c c c c c c |l}
                        Name & $p$(d) & $q = M_2/M_1$ & $M(\rm{ M}_\odot)$ & $R(\rm{R}_\odot)$ & $T_{\rm eff}$(kK) & $R/R_{\rm RL}$ & Reference \\ 
   \hline \hline
                        V382 Cyg & $1.886$ & $\num{0.726(17)}$ & $\num{26.1(4)}$ & $\num{9.4(2)}$ & $37.20^{+0.69}_{-0.72}$ & $\num{1.01}$ & A21\\
                        & & & $\num{19.0(3)}$ & $\num{8.7(2)}$ & $38.25^{+0.73}_{-0.75}$ & $\num{1.08}$ & \\
                        TU Mus & $1.387$ & $\num{0.625(9)}$ & $\num{16.7(4)}$ & $\num{7.2(5)}$ & $\num{38.7}$ & $\num{1.09}$ & P08\\
                        & & & $\num{10.4(4)}$ & $\num{5.7(5)}$ & $\num{33.2}$ & $\num{1.06}$ & \\
                        LY Aur & $4.002$ & $\num{0.550(7)}$ & $\num{25.5}$ & $\num{16.1}$ & $\num{31.0}$ & $\num{1.03}$ & Mr13 \\
                        & & & $\num{14.0}$ & $\num{12.6}$ & $\num{31.2}$ & $\num{1.03}$ & \\
                        V701 Sco & $0.762$ & $\num{0.995(2)}$ & $\num{9.78(22)}$ & $\num{4.137(316)}$ & $\num{23.5(10)}$ & $\num{1.14}$ & Y19\\
                        & & & $\num{9.74(22)}$ & $\num{4.132(312)}$ & $\num{23.44(5)}$ & $\num{1.15}$ & \\
                        CT Tau & $0.667$ & $\num{0.983(3)}$ & $\num{14.25(330)}$ & $\num{4.89(41)}$ & $\num{25.45(225)}$ & $\num{1.31}$ & Y19 \\
                        & & & $\num{14.01(340)}$ & $\num{4.89(42)}$ & $\num{25.64(12)}$ & $\num{1.33}$ & \\
                        GU Mon & $0.897$ & $\num{0.976(3)}$ & $\num{8.79(13)}$ & $\num{4.636(252)}$ & $\num{28.0(20)}$ & $\num{1.19}$ & Y19 \\
                        & & & $\num{8.58(12)}$ & $\num{4.596(246)}$ & $\num{27.82(7)}$ & $\num{1.21}$ & \\
                        XZ Cep & $5.097$ & $\num{0.50(2)}$ & $\num{18.7(13)}$ & $\num{14.2(1)}$ & $\num{28.0(10)}$ & $\num{0.85}$ & Ms17 \\
                        & & & $\num{9.3(5)}$ & $\num{14.2(1)}$ & $\num{24.0(30)}$ & $\num{1.17}$ & \\
                        LSS 3074 & $2.184$ & $\num{0.86(4)}$ & $\num{17.2(14)}$ & $\num{8.2(7)}$ & $\num{39.9(15)}$ & $\num{0.91}$ & R17 \\
                        & & & $\num{14.8(11)}$ & $\num{7.5(6)}$ & $\num{34.1(15)}$ & $\num{0.93}$ & \\
                        MY Cam & $1.175$ & $\num{0.839(27)}$ & $\num{37.7(16)}$ & $\num{7.60(10)}$ & $\num{42.0(15)}$ & $\num{1.00}$ & A21 \\
                        & & & $\num{31.6(14)}$ & $\num{7.01(9)}$ & $\num{39.0(15)}$ & $\num{1.01}$ & \\
                        V348 Car & $5.562$ & $\num{0.90(2)}$ & $\num{32(4)}$ & $\num{18.8(14)}$ & $\num{29.7(13)}$ & $\num{0.93}$ & HE85 \\
                        & & & $\num{29(4)}$ & $\num{19.3(14)}$ & $\num{26.2}$ & $\num{1.00}$ & \\
                        V729 Cyg & $6.598$ & $\num{0.290(12)}$ & $\num{31.6(29)}$ & $\num{25.6(11)}$ & $\num{28.0}$ & $\num{1.03}$ & YY14 \\
                        & & & $\num{8.8(03)}$ & $\num{14.5(10)}$ & $\num{21.26(37)}$ & $\num{1.03}$ & \\
                        BH Cen & $0.792$ & $\num{0.844(3)}$ & $\num{9.4(54)}$ & $\num{4.0(07)}$ & $\num{17.9}$ & $\num{1.09}$ & Lg84 \\
                        & & & $\num{7.9(54)}$ & $\num{3.7(7)}$ & $\num{17.43(2)}$ & $\num{109}$ & \\
                        SV Cen & $1.659$ & $\num{0.800(78)}$ & $\num{9.6}$ & $\num{7.8}$ & $\num{16}$ & $\num{1.01}$ & LS91 \\
                        & & & $\num{7.7}$ & $\num{7.3}$ & $\num{24}$ & $\num{1.08}$ & \\
            V745 Cas & 1.411 & $\num{0.57(2)}$ & $\num{18.31(51)}$& $\num{6.94(7)}$& $\num{30.00(88)}$ & 1.00 & Ç14 \\
            & & &   $\num{10.47(28)}$& $\num{5.35(5)}$ & $\num{25.60(105)}$ & 0.99 & \\
                        V606 Cen & $1.495$ & $\num{0.548(1)}$ & $\num{14.7(4)}$ & $\num{6.83(6)}$ & $\num{29.2}$ & $\num{1.01}$ & Lz99 \\
                        & & & $\num{7.96(22)}$ & $\num{5.19(5)}$ & $\num{21.77(2)}$ & $\num{1.08}$ & \\
                        HD 64315 B & $1.019$ & $\num{1.00(9)}$ & $\num{14.6(23)}$ & $\num{5.52(55)}$ & $\num{32}$ & $\num{1.10}$ & Lo17 \\
                        & & & $\num{14.6(23)}$ & $\num{5.33(52)}$ & $\num{32}$ & $\num{1.07}$ & \\
                        V4741 M31A & $1.604$ & $\num{0.924(15)}$ & $\num{18}^\dagger$ & $\num{7.2}^\dagger$ & $\num{31.6}$ & $\num{0.99}$ & Li22 \\
                        & & & $\num{16.8}^\dagger$ & $\num{6.7}^\dagger$ & $\num{27.3}$ & $\num{0.97}$ & \\
                        V1555 M31A & $0.917$ & $\num{0.974(12)}$ & $\num{23}^\dagger$ & $\num{6.7}^\dagger$ & $\num{35.1}$ & $\num{1.24}$ & Li22 \\
                        & & & $\num{22.4}^\dagger$ & $\num{6.4}^\dagger$ & $\num{34.4}$ & $\num{1.24}$ & \\
            \hline 
            VFTS 352 & $1.124$ & $\num{0.99(1)}$ & $\num{28.85(30)}$ & $\num{7.25(2)}$ & $41.45^{+1.17}_{-0.80}$ & $\num{1.01}$ & A21 \\
                & & & $\num{28.63(30)}$ & $\num{7.22(2)}$ & $44.15^{+1.1}_{-1.2}$ & $\num{1.08}$ & \\
                VFTS 066 & $1.141$ & $\num{0.52(5)}$ & $13.0^{+7.0}_{-5.0}$ & $5.8^{+0.5}_{-0.8}$ & $32.8^{+1.7}_{-1.0}$ & $\num{1.07}$ & M20 \\
                & & & $6.6^{+3.5}_{-2.8}$ & $4.4^{+0.4}_{-0.8}$ & $29.0^{+1.0}_{-1.2}$ & $\num{1.09}$ & \\
                VFTS 661 & $1.266$ & $\num{0.71(2)}$ & $27.3^{+0.9}_{-1.0}$ & $6.80^{+0.04}_{-0.01}$ & $38.4^{+0.9}_{-0.4}$ & $\num{0.94}$ & M20 \\
                & & & $19.4^{+0.6}_{-0.7}$ & $5.70^{+0.03}_{-0.01}$ & $31.8^{+1.4}_{-0.6}$ & $\num{0.92}$ & \\
                VFTS 217 & $1.855$ & $\num{0.83(1)}$ & $\num{46.8(117)}$ & $10.1^{+1.5}_{-1.2}$ & $45.0^{+0.6}_{-0.4}$ & $\num{0.91}$ & M20 \\
                & & & $\num{38.9(97)}$ & $9.4^{+1.4}_{-1.0}$ & $41.8^{+1.7}_{-0.6}$ & $\num{0.92}$ & \\
                VFTS 563 & $1.217$ & $\num{0.76(1)}$ & $26.2^{+11.9}_{-5.2}$ & $6.6^{+0.4}_{-0.7}$ & $32.4^{+1.0}_{-0.9}$ & $\num{0.95}$ & M20 \\
                & & & $20.0^{+9.1}_{-3.9}$ & $5.8^{+0.4}_{+0.6}$ & $32.4^{+1.2}_{-0.9}$ & $\num{0.95}$ & \\
                MACHO CB${}^\star$ & $1.217$ & $\num{0.64(1)}$ & $\num{41.2(12)}$ & $\num{9.56(2)}$ & $\num{50}$ & $\num{1.06}$ & O01 \\
                & & & $\num{27.0(12)}$ & $\num{7.99(5)}$ & $\num{49.5}$ & $\num{1.09}$ & \\
                SMC 108086${}^{\star \star}$ & $0.883$ & $\num{0.85(11)}$ & $\num{16.9(12)}$ & $\num{5.7(2)}$ & $36.00^{+0.69}_{-0.66}$ & $\num{1.19}$ & A21 \\
                & & & $\num{14.3(17)}$ & $\num{5.3(2)}$ & $35.20^{+0.64}_{-0.72}$ & $\num{1.19}$ & \\
   \hline
                \end{tabular}
        \tablefoot{Table after \citet{Fabry2024} (Table~C.1 and C.2). For each system, the first (second) line gives the mass, radius, effective temperature and relative overflow for the primary (secondary).
  Values marked by ${}^\dagger$ denote that $M_1$ is adopted, from which $M_2$ follows via $q$ and the radii via Kepler's third law and the fractional radii from the photometric solution.
  The horizontal line distinguishes between the systems in the Milky Way and M31 (above) and the systems in the Magellanic Clouds (below).
  \tablefoottext{${}^\star$}{ MACHO* J053441.3-693139;}
  \tablefoottext{${}^{\star \star}$} {OGLE SMC-SC10 108086}  }
                \tablebib{A21: \citet{2021A&A...651A..96A}; 
  P08: \citet{2008ApJ...681..554P}; 
  Mr13: \citet{2013A&A...559A..22M}; 
  Y19: \citet{2019AJ....157..111Y}; 
  Ms17: \citet{2017A&A...607A..82M}; 
  R17: \citet{2017A&A...601A.133R}; 
  HE85: \citet{1985MNRAS.213...75H}; 
  YY14: \citet{2014NewA...31...32Y}; 
  Lg84: \citet{1984AJ.....89..872L}; 
  LS91: \citet{1991ApJ...379..721L}; 
  Ç14: \citet{2014MNRAS.442.1560C};
  Lz99: \citet{1999A&A...345..531L}; 
  Lo17: \citet{2017A&A...606A..54L};
  Li22: \citet{2022ApJ...932...14L};
  M20: \citet{2020A&A...634A.119M};
  O01: \citet{2001MNRAS.321L..25O}              }
\end{table*}

\end{document}